\documentclass{article}

\usepackage{microtype}

\usepackage[utf8]{inputenc}
\usepackage{amsmath,amsxtra,amssymb,latexsym,amscd,amsthm,bbm,mathabx}

\usepackage{hyperref}
\usepackage{xcolor}
\usepackage{algorithm}
\usepackage{algpseudocode}
\usepackage{enumitem}
\usepackage{authblk}
\usepackage{a4wide}
\usepackage{graphicx}

\def\E{\mathbb E}
\def\R{\mathbb R}

\newtheorem{theorem}{Theorem}%[section]

\newtheorem{remark}[theorem]{Remark}

\numberwithin{equation}{section}

\title{Surrogate model for Bayesian optimal experimental design in chromatography}
\date{}
\author[1]{Jose Rodrigo Rojo-Garcia}
\author[1]{Heikki Haario}
\author[1]{Tapio Helin}
\author[2]{Tuomo Sainio}
\affil[1]{Computational Engineering, School of Engineering Science, Lappeenranta-Lahti University of Technology, Finland}
\affil[2]{Separation Science, School of Engineering Science, Lappeenranta-Lahti University of Technology, Finland}
\begin{document}

%%%%%%%%%%%%%%%%%%%%%%%%%%%%%%%%%%%%%%%%%%%%%%
\maketitle

\begin{abstract}
We applied Bayesian Optimal Experimental Design (OED) in the estimation of parameters involved in the Equilibrium Dispersive Model for chromatography with two components with the Langmuir adsorption isotherm. The coefficients estimated were Henry’s coefficients, the total absorption capacity and the number of theoretical plates, while the design variables were the
injection time and the initial concentration. The Bayesian OED algorithm is based on nested Monte Carlo estimation, which becomes computationally challenging due to the simulation time of the PDE involved in the dispersive model. This complication was relaxed by introducing a surrogate model based on Piecewise Sparse Linear Interpolation. Using the surrogate model instead of the original one reduces significantly the simulation time and approximates the solution of the PDE with high degree of accuracy. The estimation of the parameters over strategical design points provided by OED reduces the uncertainty in the estimation of parameters. Additionally, the Bayesian OED methodology shows that there is no improvement of results after certain threshold values of injection time, concentrations and number
of observation instances.\\
\textbf{Keywords: Chromatography, parameter estimation, optimal experimental design, Bayesian statistics.}
\end{abstract}

%%%%%%%%%%%%%%%%%%%%%%%%%%%%%%%%%%%%%%%%%%
\section{Introduction}
%%%%%%%%%%%%%%%%%%%%%%%%%%%%%%%%%%%%%%%%%%

Chromatography is a method to separate chemical compounds in complex mixtures by using a column packed with a selective adsorbent material. It is utilized in vastly different scales; the amounts that can be processed range from nanograms in analytical chemistry to hundreds  or thousands of tons per year in process industry. The separation is based on differences in propagation velocities of the compounds as they are eluted with a solvent that is pumped through the bed. Molecular level interactions with the adsorbent lead to a local distribution equilibrium (adsorption equilibrium) between the mobile fluid phase and the stationary adsorbent phase. Adsorption equilibrium is thus characteristic for the solution and the adsorbent but, at constant temperature, it depends on the solute concentrations only. The retention of the compounds in the column can be explained and predicted if the adsorption equilibrium isotherm is known.

The inability to predict isotherms a priori from first principles, especially for complex systems, underscores the necessity for their experimental determination. Several methods have been developed for determining the isotherms experimentally \cite{seidel2004experimental}. The accuracy and utility of the experimental data and the models used to correlate them are paramount for model-based design, a burgeoning approach in optimizing chromatographic processes. In this context, the inverse method \cite{antos2002continuous} often gains preference among researchers, particularly when an existing mathematical model of the chromatographic system is available and can be numerically solved through established computational techniques.

Methodologically, a series of chromatographic experiments are undertaken to empirically identify key parameters, both for the adsorption isotherms and for the non-idealities in fluid flow, termed dispersion. Predictions of a dynamic simulation model are matched with the experimental data that is presented as a time series of concentration values obtained at the chromatographic column's outlet. Yet, this approach is not without its challenges. Typically, a single experiment can require an hour, and real-time, high-accuracy analysis may not be possible with online detectors. Consequently, researchers may have to rely on off-line analyses that are even more time-consuming. Moreover, the errors produced by the numerical scheme used to model the chromatographic system can be significant \cite{kaczmarski2007estimation}. On the other hand, simplified isotherm models with less parameters have been observed to yield more reliable (albeit physically less meaningful) parameters with the inverse method \cite{lesko2015choice}.

Variability in model parameters introduces uncertainties that propagate through to the resultant design, posing questions about its optimality. Therefore, a rigorous exploration of how these uncertainties might be mitigated through experiment design is both intriguing and timely.
In this article, we aim to delve into these intricacies, exploring both the theoretical and practical aspects of model-based chromatographic design, with an emphasis on the role of experimentally determined isotherms and the implications of parameter uncertainty.

Mathematical description of chromatography process is typically based on systems of hyperbolic or parabolic PDEs. In this study, we focus on the Equilibrium Dispersive Model (EDM), which comprises a nonlinear parabolic equation with an adsorption term governed by the Langmuir equations. This model depends on parameters such as the number of theoretical plates, the Langmuir constants, and Henry's coefficient \cite{guiochon2006fundamentals}.
Two common approaches for the numerical solution of this problem are the finite volume methods (FVM) and the discontinuous Galerkin method (DG) \cite{javeed_thesis, javeed2013analysis, javeed2011efficient}. 
One of the FVM methods suggested in \cite{javeed2013analysis} is the Koren scheme \cite{koren1993robust}, which will be employed in our study.

Classical strategies for the model calibration by the inverse method are based on optimization algorithms (see e.g. \cite{aster2018parameter, vogel2002computational}). Here, we adopt the Bayesian approach to quantify uncertainty in our calibration process.
The Bayesian paradigm involves representing the unknown parameters as a posterior probability distribution, derived from combining prior information with the likelihood density through the Bayes' theorem \cite{gelman1995bayesian}. Posterior probability density is often intractable and advanced numerical algorithms are 
needed to characterize it such as the Markov Chain Monte Carlo (MCMC) methods \cite{robert2004monte}. The Bayes approach has been used in other contexts of chromatography to estimate parameters in mathematical models, both PDE-based or otherwise (see e.g. \cite{briskot2019prediction, he2020bayesian, heymann2023advanced, kubik2018analysis, wiczling2021application, yamamoto2021uncertainty}). Specifically, for the EDM, we focus on estimating parameters such as the Langmuir coefficients, total adsorption capacity, and the number of theoretical plates.

Mathematical modelling of chromatography process involves variables that can be controlled  when planning the experiments, the so-called design variables. The poor specification of the design variables can significantly increase the uncertainty in the posterior measure, emphasizing the role of experimental design. For example, the controllable laboratory variables can include the concentration of the injected substance and the injection time and the sampling instances.

Optimal experimental design (OED) originally emerged in the context of frequentist statistics and expanded to Bayesian approach (see e.g. \cite{chaloner1995bayesian, gelman1995bayesian, lopez2023optimal}). OED has an increasing role in chromatography in order to maximize the efficiency of the experiment, and to reduce waste and cost \cite{HIBBERT20122}. Majority of the previous literature considers experimental design through empirical response models, obtained by fitting the model to measurement data, see e.g. \cite{ATKINSON20081, pawel2015maximum}. In this work, we leverage the underlying physico-chemical model of the system to produce more precise analysis of the design problem. Consequentially, our approach requires substantially larger computational effort.
We note that, to our knowledge, OED utilizing a physico-chemical model for chromatography has been considered previously only in the frequentist context with D-OED criteria in \cite{nestlerparameter}.

Bayesian OED aims to recover a design that maximizes a given utility averaged over the Bayesian joint distribution. Various different utilities have been proposed, with the most prevalent being alphabetic criteria, notably the A- and D-optimality. The A-optimality is based on the minimization of the posterior variance \cite{alexanderian2020optimal, alexanderian2014optimal, alexanderian2016fast}. The D-optimality criteria is connected to maximizing information gain when updating prior into posterior distribution \cite{lindley1956measure, box_lucas1959}. In this work, we focus on D-optimality, also known as expected information gain, which for linear Gaussian problems reduces to minimizing the determinant of the posterior covariance matrix.

The computational challenge in Bayesian OED emerges from the need to evaluate and optimize an expected utility expressed as a high-dimensional integral. Evaluating the integrand requires evaluation of the likelihood density and, therefore, the physico-chemical model, which is expensive. Specialized numerical algorithms, typically based on Monte Carlo methods, are needed due to the nested structure of the expectation, see \cite{ryan2003estimating} and subsequent work e.g. in \cite{huan2010accelerated, huan2015numerical, huan2013simulation}. 

Minimizing computational effort in Bayesian OED is a subject of active research.
One avenue of research seeks to utilize a Gaussian approximation, such as the Laplace approximation, for the posterior distribution. However, common error bounds for this approach are expressed in the asymptotic limit of repetitive measurements or small noise. Additionally, the approximation requires the Jacobian of the forward mapping, which is not always available (see, e.g. \cite{beck2018fast, carlon2020nesterov, long2015fast, long2013fast,  long2015laplace, helin2022edge}).
Another line of work aims to approximate the forward mapping or the likelihood density using a surrogate that is fast to evaluate but maintains high accuracy. It is worth noting that the stability of such surrogate modelling was recently demonstrated in \cite{duong2023stability}. Among the various surrogate approaches, we highlight Polynomial Chaos Expansion (PCE) over sparse grids studied in \cite{xiu2010numerical} and applied in Bayesian OED in \cite{huan2010accelerated,huan2015numerical,huan2013simulation,paulson2019optimal}. Finally we can mention other approach based in a fast calculation using sparse stochastic collocation \cite{rodrigues2022tractable}.

In this study, the primary challenge with PCE-based methods lies in their sensitivity to discontinuities or regions of high gradient in the forward mapping, which can lead to artificial oscillations when smooth basis elements are utilized \cite{le2010spectral}.
Instead, we employ Piecewise Sparse Linear Interpolation (PSLI), which provides localized approximation of the forward mapping while weakening the influence of domain dimension on the approximation rate \cite{bungartz2004sparse, klimke_thesis}.

\subsection{Our contribution}
This work contributes to the estimation of parameters in the area of chromatography and the design of experiments in several ways. We consider the situation where rough estimates for the model parameters are available, e.g., after some preliminary experiments. However, the uncertainty of the model does not allow a reliable optimization of the individual sampling instants, only the concentration  of the injected material and the injection time are used as the design variables. We develop a surrogate modeling approach that reduces the computational time so that  the Bayesian optimization of experiments becomes feasible. Especially, 

\begin{itemize}
    \item  The surrogate model enables the optimization and sensitivity analyses of the design variables as well as the MCMC sampling of the model parameter posteriors. Additionally,  the  accuracy of the surrogate model is verified to be high enough for satisfactory results.
    \item We establish the minimum number of uniformly distributed sampling instants that enables  the parameters to be estimated with small uncertainty. Also, we show that after a sufficient  number of samples, using a high enough  concentration injected,  no relevant improvement in the parameter estimation can be achieved with increased experiments.   
\end{itemize}

\subsection{Structure of the paper}
The paper is organized as follows. Section 2  describes the governing equations of the EDM in chromatography, and the FVM algorithm chosen. In the same section we introduce the concept of Bayesian inverse problems, the mathematical formulation of the parameters estimation in the chromatography model, and the sampling techniques that are used for inversion. In Section 3 we formalize the definition of D-OED. After that we explain the surrogate model based on PSLI, and mathematical properties. Additionally we explain the numerical algorithms for evaluating the utility function and their order order of convergence. In Section 4 we define the numerical values of the model parameters and design variables, with which we create synthetic data for different numbers of samples in the sensor. Finally the simulation is done with the true model and the surrogate model, and the results are compared in both cases. In Section 5 we conclude the results presented in the analysis.

%%%%%%%%%%%%%%%%%%%%%%%%%%%%%%%%%%%%%%%%%%%
\section{Mathematical preliminaries}
%%%%%%%%%%%%%%%%%%%%%%%%%%%%%%%%%%%%%%%%%%%

\subsection{Dispersive model for chromatography}

In a chromatography process, dispersive models are particularly valuable for describing the behavior of components within complex mixtures as they travel through a chromatographic column. In them, the system is characterized by a set of components $c_i$, where $i$ ranges from 1 to $N_c$, representing the different species or compounds of interest. These components are governed by a set of nonlinear hyperbolic partial differential equations (PDEs). Each component $c_i$ is described by the following governing equation
\begin{equation}
\frac{\partial c_i}{\partial t} + F \frac{\partial q_i}{\partial t} + u \frac{\partial c_i}{\partial z} = D_{\text{app}} \frac{\partial^2 c_i}{\partial z^2}, \quad (z,t) \in (0,L) \times (0,T),
\label{Eq1}
\end{equation}
where $F$ is the ratio of volumetric fractions of stationary and mobile phases, $q_i$ models the isotherms of the system, $u$ is the linear velocity of the mobile phase, $D_{app}$ is the axial dispersion coefficient, $L$ is the length of the column and $T$ is the total time simulation.
The equation \eqref{Eq1} is subject to the zero initial conditions
\begin{equation}
c_i(z,0) = 0, \quad q_i(z,0) = 0, \quad z \in (0,L).
\label{Eq2}
\end{equation}
In particular, the latter initial condition indicates that at the column inlet, no adsorption has occurred. A common model, connecting the isotherms $q_i$ to the concentrations $c_i$, is given by the Langmuir equation
\begin{equation}
q_i(z,t)=\frac{Q_{s} b_i c_i(z,t)}{1+\sum_{l=1}^{N_{c}} b_{l} c_{l}(z,t)}, \quad (z,t) \in (0,L) \times (0,T),
\label{Eq6}
\end{equation}
where $q_i$ represents the maximum adsorption capacity, $b_i$ is the Langmuir constant for component $i$, and $Q_s$ is the total adsorption capacity of the stationary phase. 

The inflow and outflow of fluids in the boundaries are specified by Danckwerts conditions \cite{guiochon2006fundamentals}.
\begin{align*}
c_i(0,t) &= c_{i,0}(t) + \frac{D_i}{u} \frac{\partial c_i}{\partial z} (0,t), \quad i=1,2,\ldots, N_{c}, \\
\frac{\partial c_i}{\partial z}(L, t) &= 0.
\end{align*}

Notice that at the column's outlet ($z = L$), the spatial gradient of each component is set to zero, ensuring no mass transfer at the exit.
In what follows, we consider injections $c_{i,0}(t)$ satisfying
\begin{equation}
c_{i,0}(t)=
\begin{cases}
c_i^{\text{Feed}} & \text{for } t \leq t^{\text{inj}}, \\
0 & \text{for } t > t^{\text{inj}},
\end{cases} \quad i = 1, \ldots, N_{c}.
\label{Eq5}
\end{equation}

A common assumption is the dependency between the dispersion parameter $D_{app}$ and the velocity $u$, and a common choice is the linear relation \cite{guiochon2006fundamentals}
\begin{equation*}
D_{\mathrm{app}}=\frac{L u}{2 N_{tp}},
\end{equation*}
here the parameter $N_{tp}$ refers to theoretical plates formed during the process.

For numerical purposes, we rescale spatial and time variables according to
\begin{equation*}
    t = \frac Lu\tau, \quad \tau\in [0,\Upsilon) \quad \text{and} \quad
    z = Ly, \quad y\in [0,1], 
\end{equation*}
where $\Upsilon = \frac{uT}L$ is a dimensionless time. Then, the problem is transformed to
\begin{equation}
\frac{\partial c_i}{\partial \tau}+F \frac{\partial q_i}{\partial \tau}+\frac{\partial c_i}{\partial y}=\frac{1}{2 N_{t p}} \frac{\partial^{2} c_i}{\partial y^{2}},
\label{EDM1}
\end{equation}
with initial conditions
\begin{equation}
c_i(y,0)=0, \quad q_i(y,0)=0,
\label{EDM2}
\end{equation}
and boundary conditions
\begin{equation}
\left.c_i\right|_{y=0}=c_{i,0}+\left.\frac{1}{2 N_{t p}} \frac{\partial c_i}{\partial y}\right|_{y=0}, \quad \frac{\partial c_i(1, \tau)}{\partial y}=0,
\label{EDM3}
\end{equation}
and where
\begin{equation}
c_{i,0}(\tau)=\left\{\begin{array}{ll}
c_i^{\text {Feed }} & \text { for } \tau \leq \tau^{\mathrm{inj}} \\
0 & \text { for } \tau>\tau^{\mathrm{inj}}
\end{array}\right.
\label{EDM4}
\end{equation}
With this change of variables, the transformed model depends on $u$ only through the new time variable $\tau$ which simplifies the structure of the model.

\begin{remark}
\label{remark1}
The regularity of the solutions  the  PDE system  depends on the regularity of the boundary conditions and the initial condition: for regular enough conditions the solution is differentiable.  More mathematically (see \cite{ladyzenskaja1988linear, lieberman2005second, zhang2017adjoint}), fixing the model parameters and assuming $c_{i}(0,\cdot)\in (L^{2}(0,L))^{N_{c}}$ and $c_{i,0}\in (L^{2}(0,T))^{N_{c}}$, there exist a unique and stable solution in $c_{i}\in (W_{2}^{2,1}((0,L)\times (0,T)))^{N_{c}}$.    
\end{remark}

\subsection{Numerical implementation by the Koren scheme}

The Koren scheme is a FVM method based on a piecewise-polynomial interpolation for the flux, and a flux limiter proposed by Sweby \cite{sweby1984high}. Consider regular grid $y_m$, $m=0,...,N_{t} + 1$ on a unit interval with $y_0 = 0$ and $y_{N_{t} + 1} = 1$. We denote the mid-points of the grid by $y_{m+1/2}$ for $m=0,...,N_t$, and the stepsize by $\Delta y$, respectively.

For what follows, let us also denote $\bar c = (c_i)_{i=1}^{N_c} : (0,1) \times (0,\Upsilon) \to \R^{N_c}$, and the flux as $f(\bar c) = \bar c.$ By chain rule, we have
\begin{equation*}
    \frac{\partial q_i}{\partial \tau}(y,\tau)
    = \sum_{j=1}^{N_c} \frac{d q_i}{d c_j}(y,\tau) \frac{\partial c_j}{\partial \tau}(y,\tau) = (\nabla q_i)(y,\tau) \cdot \frac{\partial \bar c}{\partial \tau}(y,\tau)
\end{equation*}
with the convention $(\nabla q_i) = (\frac{d q_i}{d c_j})_{j=1}^{N_c} : (0,1) \times (0,\Upsilon) \to \R^{N_c}$. In consequence, the vectorized form of the system (\ref{EDM1}) is 
\begin{equation*}
    (I+FQ)\frac{\partial \bar c}{\partial \tau} = - \frac{\partial \bar c}{\partial y} + \frac{1}{2N_{tp}} \frac{\partial^2 \bar c}{\partial y^2},
\end{equation*}
where $I$ stands for an $N_c\times N_c$ identity matrix and $Q$ is composed of the elements
\begin{equation*}
    Q_{ij} = \frac{dq_i}{dc_j} : (0,1)\times (0,\Upsilon) \to \R, \quad 1\leq i,j \leq N_c.
\end{equation*}

For $m =1,2,\cdots,N_t$, the Koren scheme is given by
\begin{multline}
\frac{d \bar c}{d \tau}(y_m,\tau) \\
=-\left(I+F Q(y_m,\tau)\right)^{-1}\left[\frac{f_{m+\frac{1}{2}}(\tau)-f_{m-\frac{1}{2}}(\tau)}{\Delta y}-\frac{1}{2N_{tp}\Delta y}\left(\frac{\partial \bar c}{\partial y}(y_{m+\frac{1}{2}},\tau)-\frac{\partial \bar c}{\partial y}(y_{m-\frac{1}{2}},\tau)\right)\right],
\label{Eq7}
\end{multline}
and
\begin{equation*}
f_{m+\frac{1}{2}}=f_{m}+\frac{1}{2} \phi\left(r_{m+\frac{1}{2}}\right)\left(f_{m}-f_{m-1}\right).
\end{equation*}
Above, the term $\phi$ is the flux limiter and it is defined as 
\begin{equation*}
\phi\left(r_{m+\frac{1}{2}}\right)=\max \left(0, \min \left(2 r_{m+\frac{1}{2}}, \min \left(\frac{1}{3}+\frac{2}{3} r_{m+\frac{1}{2}}, 2\right)\right)\right),
\end{equation*}
where 
\begin{equation*}
r_{m+\frac{1}{2}}=\frac{f_{m+1}-f_{m}+\eta}{f_{m}-f_{m-1}+\eta},
\end{equation*}
represents the ratio between consecutive fluxes with some small $\eta>0$ to avoid numerical singularities. In the cases $m = 1$ and $N_t$, the fluxes are calculated as follows
\begin{equation*}
f_{\frac{1}{2}}(\tau)=\bar c(0, \tau), \quad f_{\frac{3}{2}}(\tau)=\bar c(y_1,\tau), \quad f_{N_{t}+\frac{1}{2}}(\tau)=\bar c(y_{N_t},\tau).
\end{equation*}

Furthermore, for approximate the partial derivatives we utilize the central scheme
\begin{equation*}
\frac{\partial \bar c}{\partial y}(y_{m+\frac{1}{2}},\tau) \approx \frac{\bar c(y_{m + 1},\tau)- \bar c(y_m,\tau)}{\Delta y}.
\end{equation*}
Consider the local truncation error defined by
\begin{multline*}
\bar E(\tau) := \frac{d \bar c}{d \tau}(y_m,\tau) \\
+\left(I+F Q(y_m,\tau)\right)^{-1}\left[\frac{f_{m+\frac{1}{2}}(\tau)-f_{m-\frac{1}{2}}(\tau)}{\Delta y}-\frac{1}{2N_{tp}\Delta y}\left(\left(\frac{\partial \bar c}{\partial y}\right)_{m+\frac{1}{2}}-\left(\frac{\partial \bar c}{\partial y}\right)_{m-\frac{1}{2}}\right)\right].
\end{multline*}
The Koren scheme has spatial consistency order 2, i.e., $\Vert \bar E(\tau) \Vert \leq \mathcal{O}(|\Delta y|^{2})$ (see e.g. \cite{javeed_thesis, javeed2011efficient, koren1993robust}). Following the methodology in \cite{javeed_thesis}, we use Runge-Kutta 4 (RK4) for temporal discretization since it is well known that RK4 has consistency order 5.

%%%%%%%%%%%%%%%%%%%%%%%%%%%%%%%%%%%%%%%%%%%%%%%%%%
\subsection{Bayesian inference}
%%%%%%%%%%%%%%%%%%%%%%%%%%%%%%%%%%%%%%%%%%%%%%%%%%

Let us now  formalize our experiment. The solution $c_i$, $i=1,...,N_c$ of the system \eqref{EDM1}-\eqref{EDM4} is specified by the values of parameters $F$, $N_{tp}$, $b_1,...,b_{N_c}$, $Q_s$, $t^{\mathrm{inj}}$, $c_1^{\mathrm{Feed}}, ..., c_{N_c}^{\mathrm{Feed}}$. For what follows, let ${\mathcal P}$ stand for the Cartesian product of all individual parameter domains. We next decompose ${\mathcal P}$ into three components ${\mathcal P} = \Theta \times \mathcal{D} \times \mathcal{E}$, where $\Theta$, $\mathcal{D}$ and $\mathcal{E}$ denote the sets of parameters of interest, design parameters in our experiment and fixed variables, respectively. The isotherm parameters $b_j$, $Q_s$ and $N_{tp}$  are the components of $\Theta$ that need to be estimated. The design variables  $\mathcal{D}$  are the feed concentrations $c_j^{\mathrm{Feed}}$ and the injection time $t^{\mathrm{inj}}$.  And finally, the only fixed variable in the model is $F$.
Below, the process of estimating these parameters of interest is referred to as the inverse problem.  We define a mapping from the model parameters and design variables to the model solutions as  $G:\Theta \times \mathcal{D} \rightarrow (W_{2}^{2,1}((0,1)\times (0,\Upsilon)))^{N_c}$ as the unique solution of the system \eqref{EDM1}-\eqref{EDM4} specified by Remark \ref{remark1}.
In addition, let ${\mathcal O}(\cdot) : (W_{2}^{2,1}((0,1)\times (0,\Upsilon)))^{N_c} \rightarrow \R^K$ denote the observation operator evaluating the solution at a grid of $K$ time points on $(0,1)\times (0,\Upsilon)$. We remark that in chromatography we only do measurements at the outlet $y = 1$.

In the experiment, we wish to determine the unknown parameters $\theta \in \Theta$ from the noisy observations
\begin{equation}
	\label{eq:model_equation}
	{\bf c} = \mathcal{O}(G(\theta; d)) + \eta = \mathcal{G}(\theta; d) + \eta,
\end{equation}
where ${\bf c}, \eta \in \R^K$ represent vectors of point evaluations of the concentrations and noise, respectively, and we write $\mathcal{G} = \mathcal{O}\circ G : \Theta \times \mathcal D \to \R^K$. The realization of the noise process is assumed to be drawn from the Gaussian distribution ${\mathcal N}(0,\sigma^2 I)$, i.e., each point evaluation is independent with the same noise level $\sigma$. The value of $\sigma$ is supposed to be known, as estimated by fitted residuals. 

Due to the presence of noise in the data, there is inherent uncertainty associated with any estimator of $\theta$. The Bayesian inference provides a paradigm to assess this uncertainty by updating any prior distribution of $\theta$ into a posterior distribution. We assume here Gaussian measurement noise. The parameter uncertainty  is then formalized by the next theorem.

\begin{theorem}[Bayes' theorem]
\label{thm:Bayes}
Let $p_0(\theta)$ denote the prior probability density of $\theta \in \Theta$. Then the posterior density function $p(\theta | {\bf c};d)$ (the density of $\theta$ given ${\bf c} \in \R^K$) corresponding to the measurement \eqref{eq:model_equation} is 
given by
\begin{equation*}
\end{equation*}
where
\begin{equation*}
	Z(\mathbf{c},d) = \int_\Theta \exp\left(-\frac{\Vert \mathcal{G}(\theta; d) - \mathbf{c}^2\Vert}{2\sigma^2}\right) p_0(\theta)d\theta.
\end{equation*}
\end{theorem}

To explore the posterior distribution, Markov Chain Monte Carlo (MCMC) methods are commonly employed, as they can generate samples from the posterior distribution without the need for the normalization constant, as discussed in \cite{robert2004monte}. Producing a sample size that accurately represents the final distribution may demand substantial computational effort. Various adaptive techniques have been developed, as discussed in \cite{christen2010general, haario1999adaptive, haario2001adaptive}. One pragmatic and efficient approach is the DRAM algorithm, as detailed in \cite{haario2006dram}, which is supported by a MATLAB toolbox and has demonstrated satisfactory results.

%%%%%%%%%%%%%%%%%%%%%%%%%%%%%%%%%%%%%%%%%%%%%%%%%%
\section{Optimal experimental design}
%%%%%%%%%%%%%%%%%%%%%%%%%%%%%%%%%%%%%%%%%%%%%%%%%%

In this section, we formulate Bayesian approach to experimental design. Here we assume that we have only two components and the initial injected concentration is equal for both, i.e., $c_{1}^{Feed} = c_{2}^{Feed} = c^{Feed}$. Then, the uncertainty variables are $\theta = (b_{1},b_{2},Q_{s},N_{tp})$, the design variables are $d = (\tau^{inj}, c^{Feed})$ and the measurements are given by a vector of size $K = 2N_s$ with evaluations of $c_1$ and $c_2$ in the outlet $y = 1$ and at times $\tau_1, \tau_2, \cdots, \tau_{N_s}$ as
\begin{equation*}
    {\bf c} = (c_{1}(1,\tau_1),c_{1}(1,\tau_2),\cdots,c_{1}(1,\tau_{N_s}),c_{2}(1,\tau_1),c_{2}(1,\tau_2),\cdots,c_{2}(1,\tau_{N_s})).
\end{equation*}

Bayesian optimal experimental design aims to maximize the expected utility associated with a given design variable \cite{chaloner1995bayesian}. In more precise mathematical notation, the objective is to maximize the utility function
\begin{equation*}
	U(d) = \int_{\R^K\times \Theta} u(\theta, {\bf c}; d) p(\theta,{\bf c};d)d\theta d{\bf c},
\end{equation*}
where $p(\theta,{\bf c};d)$ is the Bayesian joint density function of $\theta$ and ${\bf c}$ on $\R^K \times \Theta$ and $u$ is the information gain. 

Here, we consider the D-OED criteria based on the Kullback--Leibler divergence (KL) between the posterior and prior measures, i.e
\begin{equation*}
	u(\theta, {\bf c}; d) = D_{KL}(p(\cdot \mid {\bf c}; d) \; \| \; p_0),
\end{equation*}
and the expected utility function is then reduced as
\begin{equation}
	\label{eq:expected_IG}
	U(d) = \E^{{\bf c}} D_{KL}(p(\cdot \mid {\bf c}; d) \; \| \; p_0)
	= \iint_{\R^K \times \Theta} p(\theta | {\bf c} ;d) \log \left(\dfrac{p(\theta | {\bf c};d)}{p_0(\theta)}  \right) d\theta p({\bf c} ; d) d{\bf c},
\end{equation}
where $p({\bf c} ; d)$ is the marginal density of the measurement ${\bf c}$. 
This quantity indicates how similar are posterior and prior measures in the design node $d$, and a big value implies a more informative posterior distribution in that node. 

A main challenge of Bayesian experimental design is the numerical approximation of the above double integral. 
For linear inverse problems with Gaussian prior distribution, the expected information gain \eqref{eq:expected_IG} can be evaluated in closed form and it is reduced to the computation of the log-determinant of the posterior covariance matrix. In the inverse problem literature, this property has been widely utilized in various contexts; see, e.g., \cite{alexanderian2020optimal, alexanderian2018efficient, solonen2012simulation}.

For nonlinear forward mappings that do not have sufficient linearization properties, such as our problem here, the integral in \eqref{eq:expected_IG} needs to be evaluated approximatively. While Monte Carlo-based algorithms are a natural choice, they tend to require high computational effort 
due to the inherent nested sampling task, which we discuss more below.
Before considering evaluation, let us note that the computational cost is dominated by the effort needed to evaluate the forward mapping. Well-designed surrogate models can reduce this cost.

\subsection{Double-loop Monte Carlo integration}

The expected utility in \eqref{eq:expected_IG} can be rephrased as
\begin{equation}
\label{eq:expected_IG_2}
U(d) =  \int_{\R^K \times \Theta}\{\log [p({\bf c} | \theta; d)]-\log [p({\bf c} ; d)]\} p({\bf c}, \theta; d) d \theta d {\bf c},
\end{equation}
where $p({\bf c} ; d)$ refers to the marginal density of the observation, i.e., the evidence.
The computational challenge in estimating \eqref{eq:expected_IG_2} is that the integrand function is intractable (excluding special cases) and cannot be directly approximated by conventional Monte Carlo methods.
In literature, the traditional approach is to apply double-loop (i.e. nested) Monte Carlo integration as follows: first, an ensemble $\{(\theta^k, {\bf c}^k)\}_{k=1}^M$ is generated from the joint distribution. Second, noting the identity
\begin{equation*}
	p\left({\bf c}^k ; d\right)=\int_{\Theta} p\left({\bf c}^k | \theta; d\right) p_{0}(\theta) d \theta,
\end{equation*}
the evidence density $p({\bf c}^k ; d)$ is approximated using a prior ensemble $\{\theta^{j,k}\}_{j=1}^J$ for each $k=1,...,M$.
This leads to the nested MC approximation 
\begin{equation*}
%\label{eq:expected_IG_MC1}
U_{M}(d) = \frac{1}{M} \sum_{k=1}^{M} \log \left[p\left({\bf c}^{k} | \theta^{k}; d\right)\right]- \frac 1{JM} \sum_{k=1}^M \sum_{j=1}^J \log \left[p\left({\bf c}^{k} | \theta^{j,k} ; d\right)\right].
\end{equation*}
Notice that the estimator $U_M$ is biased positively, i.e. $\mathbb{E}[U_{M}(d) - U(d)]\geq 0$ and the asymptotic mean squared error (MSE) is of the order $\mathbb{E} |U_{M}(d)-U(d)|^2 = \mathcal{O}(1/M + 1/J^2)$ (see \cite{ryan2003estimating}). By scaling $J\propto \sqrt M$, we obtain MSE error ${\mathcal O}(1/M)$ with the computational cost of $M^{3/2}$ number of likelihood evaluations and, therefore, evaluations of the forward operator ${\mathcal G}$.
The line of work \cite{huan2010accelerated, huan2015numerical, huan2013simulation} suggest replacing the nested ensemble $\{\theta^{j,k}\}_{j=1}^J$ for each $k=1,...,M$, by the original prior ensemble $\{\theta^k\}_{k=1}^M$, reducing the computational effort.
The resulting method is summarized in Algorithm \ref{alg:cap}.

\begin{algorithm}
\caption{Accelerated double-loop Monte Carlo estimation of expected utility following \cite{huan2010accelerated}}\label{alg:cap}
\begin{algorithmic}
\Require Ensemble $\lbrace \theta_{k} \rbrace_{k=1}^{M} \sim p_0(\theta)$ i.i.d. Design variable $d$.   
\Ensure $U(d)$
\State $U(d) \gets 0$
\For{$k = 1:M$}
\State $\widehat{{\bf c}}^{k} \gets {\mathcal G}(\theta^{k}; d)$ \Comment{Map all elements of the ensemble.}
\EndFor

\For{$k = 1:M$} \Comment{Outer loop}
%\State $\widehat{{\bf c}}^{k} \gets \widehat{C}(k,:)$
\State ${\bf c}^{k} \sim \mathcal{N}(\widehat{{\bf c}}^{k},\sigma^{2}I)$    %\Comment{Generate a sample from likelihood.}
\State $\log \left[p\left({\bf c}^{k} \mid \ \theta^{k}, d\right)\right] \gets -\dfrac{\Vert {\bf c}^{k} - \widehat{{\bf c}}^{k} \Vert_{2}^{2}}{2\sigma^{2}}$    

\State $p\left({\bf c}^{k} \mid d\right) \gets 0$  
\For{$j = 1:M$} \Comment{Inner loop}
%\State $\widehat{{\bf c}}^{j} \gets \widehat{C}(j,:)$
\State $p\left({\bf c}^{k} \mid d\right) \gets p\left({\bf c}^{k} \mid d\right) - \frac 1M \exp\left( -\dfrac{\Vert {\bf c}^{k} - \widehat{{\bf c}}^{j} \Vert_{2}^{2}}{2\sigma^{2}}\right)$
\EndFor
\State $U(d) \gets U(d) + \frac 1M\left(\log \left[p\left({\bf c}^{k} \mid \ \theta^{k}, d\right)\right] - \log \left[p\left({\bf c}^{k} \mid d\right) \right]\right)$  

\EndFor

\end{algorithmic}
\end{algorithm}

%%%%%%%%%%%%%%%%%%%%%%%%%%%%%%%%%%%%%%%%%%%%%%%%%%
\subsection{Surrogated model based in Piecewise Linear Interpolation}
%%%%%%%%%%%%%%%%%%%%%%%%%%%%%%%%%%%%%%%%%%%%%%%%%%

In our context, the main computational effort associated to the double-loop MC Algorithm \ref{alg:cap} arises from the repeated evaluations of the likelihood density and, therefore, of the forward operator ${\mathcal G}$. Consequently, it is tempting to replace ${\mathcal G}$ with a surrogate that offers both accurate approximation and reduced computational cost.  
For computational purposes, we consider the approximative observational model \eqref{eq:model_equation} by an approximation
\begin{equation*}	
%	\label{eq:surrogate_model}
	{\bf c}_{N} = {\mathcal G}_N(\theta; d) + \eta
\end{equation*}
with a surrogate mapping ${\mathcal G}_N: \Theta \times \mathcal{D} \rightarrow \R^K$. 
The parameter $N$ identifies the approximation rate, which here will correspond to the discretization grid.
In \cite{duong2023stability}, it was demonstrated that the $U(d)$ is stable w.r.t. perturbations in the likelihood. In particular, it was shown that the induced error in EIG evaluation is bounded by the squared approximation error ${\mathcal G}(\theta; d) - {\mathcal G}_N(\theta; d)$ averaged over the prior.

%If ${\mathcal G}_N$ is continuous and bounded, the inverse problem is well-posed.

In this work, we utilize piecewise sparse linear interpolation (PSLI) \cite{bungartz2004sparse}, which is well-suited for surrogate modelling of mappings with limited smoothness. For their rigorous construction, let us begin by introducing
a one-dimensional piecewise linear interpolant $U[f]$ of a function $f:[-1,1]\rightarrow \mathbb{R}$. This interpolant is defined over a set of nodes $-1 = x_{0} < x_{1} < x_{2} < \cdots < x_{m} = 1$ and a nodal basis $\lbrace a_{j} \rbrace_{j=1}^{m}$ as follows:
\begin{equation*}
%    \label{eq:unidim_interpolant}
   U[f](x) = \sum_{j=0}^{m}f(x_j)a_{j}(x), \quad x\in[-1,1], 
\end{equation*}
where the basis is given by the hat functions
$$
   a_j(x)= \begin{cases}1-\frac{n\left|x-x_j\right|}{2}, & \text { if }\left|x-x_j\right|< \frac{2}{m} \\ 0, & \text { otherwise. }\end{cases} 
$$
In what follows, we utilize the Clenshaw--Curtis quadrature rule defined by nodes $x_j = -\cos(j \pi/ m)$. 
Generalizing to higher dimensions, we denote $U^{i}$ for the one-dimensional interpolant acting on the $i^{th}$ coordinate. The multivariate interpolant on $[-1,1]^r$ is then constructed as a tensor product
\begin{equation}
    \label{eq:multivariateU}
   (U^{i_1} \otimes ... \otimes U^{i_r})[f](x^1,\cdots,x^r) = \sum_{j_1=0}^{m_1}\sum_{j_2=0}^{m_2}\cdots \sum_{j_r=0}^{m_r} f(x_{j_1}^1,\cdots,x_{j_r}^r)(a_{j_1}(x^1) \cdots a_{j_r}(x^r)),  
\end{equation}
where $(x^1,x^2,\cdots,x^r) \in [-1,1]^r$.

The drawback of \eqref{eq:multivariateU} is the exponentially growing number of function evaluations w.r.t. the dimension $r$ when $m_1 = ... = m_r$. 
Sparse grid methods \cite{bungartz2004sparse} provide a remedy for this curse of dimension by limiting the sum in \eqref{eq:multivariateU} to a carefully designed sparse subset of the grid points without sacrificing too much accuracy. The Smolyak formulas \cite{smolyak1963quadrature} construct
the sparse grid leading to a hierarchical basis $\lbrace a_{j_{k}}  \rbrace_{k=1}^{r}$ and preserving the interpolation properties of $r=1$ when extended to higher dimensions.

\begin{figure}[htp]
%\centering\includegraphics[scale = 0.4]{Plots/CC_3.pdf}
\centering\includegraphics[scale = 0.43]{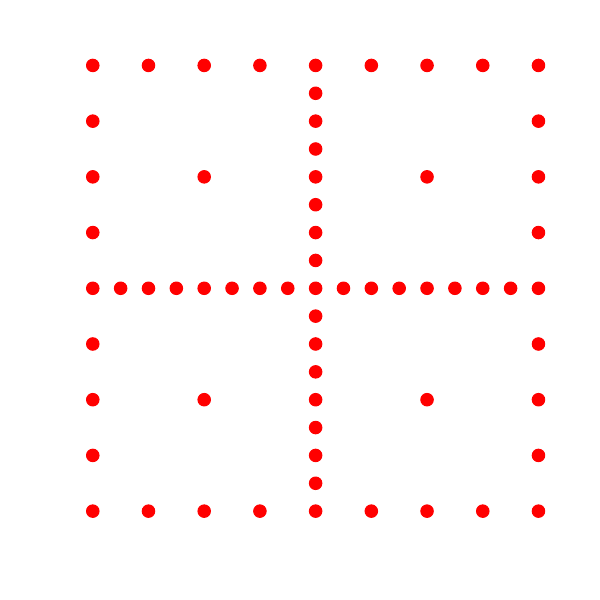}
\centering\includegraphics[scale = 0.43]{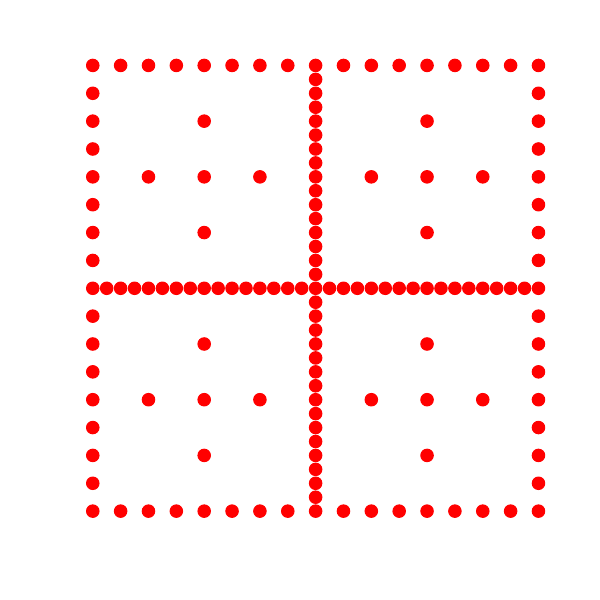}
\centering\includegraphics[scale = 0.43]{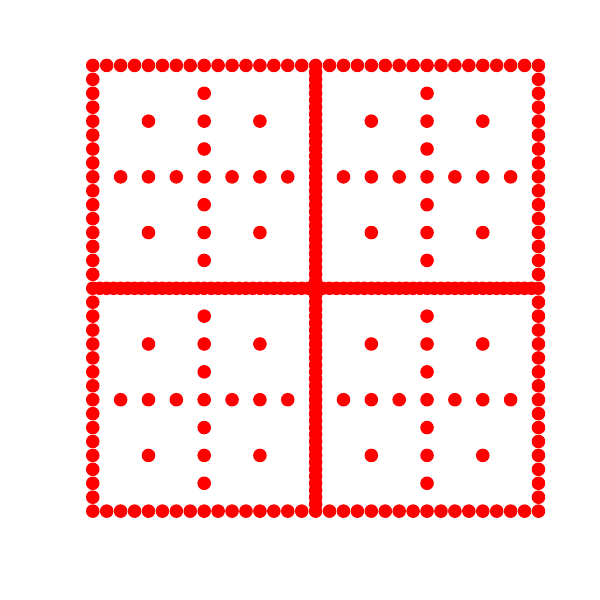}

\caption{Smolyak-Sparse Grids in 2D. The number of nodes are $N = 65, 145, 321$ for left, center and right respectively.}
\label{Plot_Smolyak_Grid}
\end{figure}

With the convention $U^0 = 0$, the Smolyak formula is defined as follows: given $q \geq r$ we set
\begin{eqnarray}
S_{q,r}& =& \sum_{|\mathbf{i}| \leq q} \left((U^{i_1}-U^{i_1-1}) \otimes \cdots \otimes (U^{i_r}-U^{i_r-1})\right) \nonumber \\
& = & \sum_{q-r+1 \leq \vert \mathbf{i} \vert \leq q}(-1)^{r-|\mathbf{i}|} \cdot\left(\begin{array}{c}
r-1 \\
q-|\mathbf{i}|
\end{array}\right) \cdot\left(U^{i_1} \otimes \cdots \otimes U^{i_r}\right),
\label{Smolyak_interp}
\end{eqnarray}
where ${\bf i} = (i_1, ..., i_r)\in {\mathbb N}^r$ and $|{\bf i}| = i_1 + ... + i_r$. For the second identity, see e.g. \cite[Lemma 1]{wasilkowski1995explicit}.
The resulting grid consists of $N(q,l)\leq 2^{q}\binom{q-1}{l-1}$ points (see e.g. \cite{novak1996high}); the arrangement is illustrated on a two-dimensional domain in Figure \ref{Plot_Smolyak_Grid}.

The interpolant $S_{q,r}$ has well-established approximation properties. For the characretization, we introduce the function space
(see \cite{bungartz1998finite} and \cite{novak1996high}) 
\begin{equation*}
F_{r}^{k} = \begin{cases}
C^{k}([-1,1]) & \text { if } r =1 \\ 
\left\{f:[-1,1]^r \rightarrow \mathbb{R} \mid D^\alpha f \text { continuous if } \alpha_i \leq k \text { for all } i\right\} & \text { if } r > 1\end{cases}
\end{equation*}

%\begin{theorem}[\cite{novak1996high}]
\begin{theorem}[{\cite[Remark 10]{barthelmann2000high}}]
\label{convergence_PSLI}
Given $f\in F_r^{k}$, $k\in\{1,2\}$, the interpolant $S_{q,r}$ satisfies
\begin{equation*}
\Vert f-S_{q,r}f \Vert_{\infty} \leq c_{r,k} N^{-k} (\log N)^{(k+1)(r-1)} \Vert f\Vert_{C^k},
\end{equation*}
where $N=N(q,r)$ is the number of grid points evaluated in $S_{q,r}$.
\end{theorem}

We note that piecewise linear interpolation is not order optimal for $k>2$ and, therefore, the rate in Theorem \ref{convergence_PSLI} does not extend beyond $k=2$, see \cite[Remark 10]{barthelmann2000high}.
An efficient algorithmic implementation is discussed in \cite{klimke_thesis, klimke2006efficient}, and a free toolbox is available \cite{klimke2008sparse, klimke2005algorithm}.
 Moreover, Theorem \ref{convergence_PSLI} does not guarantee convergence for continuous functions that are not continuously differentiable. Nonetheless, continuity is typically the minimum requirement for interpolation, and numerical experiments in \cite{klimke_thesis, novak1996high} show convergence for continuous benchmark functions. Finally, we define our surrogate model over the parameter space $\Theta$ by constructing
\begin{equation*}
    {\mathcal G}_N(\cdot; d_j) = (S_{q,r}{\mathcal G}_k(\cdot; d_j))_{k=1}^K,
\end{equation*}
with ${\mathcal G} = ({\mathcal G}_k)_{k=1}^K$, on a grid of points $d_j \in {\mathcal D}$, where $N = N(q,r)$. The full mapping for all $d\in {\mathcal D}$ is obtained by linear interpolation.
We note that, while theoretically establishing smoothness of the forward mapping ${\mathcal G}$ is outside the scope of this work, no issues were encountered in our numerical simulations.

%%%%%%%%%%%%%%%%%%%%%%%%%%%%%%%%%%%%%%%%%%%%%%%%%%

%%%%%%%%%%%%%%%%%%%%%%%%%%%%%%%%%%%%%%%%%%%%%%%%%%
\section{Numerical simulations}
%%%%%%%%%%%%%%%%%%%%%%%%%%%%%%%%%%%%%%%%%%%%%%%%%%,

%%%%%%%%%%%%%%%%%%%%%%%%%%%%%%%%%%%%%%%%%%%%%%%%%%
\subsection{Problem formulation}
%%%%%%%%%%%%%%%%%%%%%%%%%%%%%%%%%%%%%%%%%%%%%%%%%%

In our numerical simulations, we assume that our measurements are obtained on an equidistant temporal grid with $N_s$ nodes and the measurement data is contaminated with a normally distributed noise with standard deviation $\sigma$.

We utilize a conservative prior model composed of independent uniform distributions for every component of the parameter $\theta$. That is, for every parameter $\theta_i$, we define a uniform prior in each interval $\left[\alpha_i, \beta_i\right] \subset \R_+$, i.e., 
$\theta_i \sim \mathcal{U}\left(\left[\alpha_i, \beta_i\right]\right)$. Furthermore, we can restrict the domain to the Cartesian product of the intervals
$\Theta = \times_{i=1}^{4}\left[\alpha_{i},\beta_{i}\right]$. We assume that these intervals can be reliably specified from independent information, e.g. a rough preliminary experiment. We note that a similar methodology was utilized in \cite{wiczling2016pharmacokinetics}. 

As earlier, our design variables consist of the injection time and concentration. Their range is limited due to the experimental conditions at the laboratory. 
In consequence, we specify the design domain as the Cartesian product $
\mathcal{D} = \left[\tau_{0}^{\text{inj}},\tau_{F}^{\text{inj}}\right] \times \left[c_{0}^{\text{Feed}},c_{F}^{\text{Feed}}\right]$.  

\subsection{Simulations}

For this study, we exclusively utilized synthetic data, which we divided into three experiments. In each experiment, measurement data was generated on an equidistant grid with
$N_{s} = 8, 15$, and $20$ temporal nodes within the time interval $[0.5,9.5]$. 

In the spirit of \cite{bozorgnia2020numerical}, we specified the true data-generating parameters as $b_{1} = 0.05$ [L/mol] , $b_{2} = 0.10$ [L/mol], $Q_{s} = 10$ [mol/L]. Moreover, we set $N_{tp} = 70$. The standard deviation of noise was $\sigma = 0.05$ [mol/L], and other parameters satisfied $\Upsilon = 10$ and the constant $F = 1.5$. The intervals of the prior distribution and for the design space are shown in the tables \ref{table:1} and \ref{table:2}, respectively.

\begin{table}[h!]
\centering
\begin{tabular}{||c c c c c||} 
 \hline
 $Parameter$ & Lower Bound & Upper Bound & Real parameter & Units \\ [0.5ex] 
 \hline\hline
  $b_{1}$ & 0.02 & 0.08 & 0.05 & L/mol\\ 
 $b_{2}$ & 0.03 & 0.17 & 0.1 & L/mol\\
 $Q_{s}$ & 8 & 11 & 10 & mol/L\\
 $N_{tp}$ & 50 & 180 & 70 & -\\ [1ex] 
 \hline
\end{tabular}
\caption{Uncertainty Parameters.}
\label{table:1}
\end{table}

\begin{table}[h!]
\centering
\begin{tabular}{||c c c c||} 
 \hline
 $Parameter$ & Lower Bound & Upper Bound & Units\\ [0.5ex] 
 \hline\hline
   $\tau^{inj}$ & 0.05 & 3 & -\\ 
 $c^{Feed}$ & 1 & 15 & mol/L\\ [1ex] 
 \hline
\end{tabular}
\caption{Design Parameters.}
\label{table:2}
\end{table}

\begin{figure}[htp]
\centering\includegraphics[scale = 1]{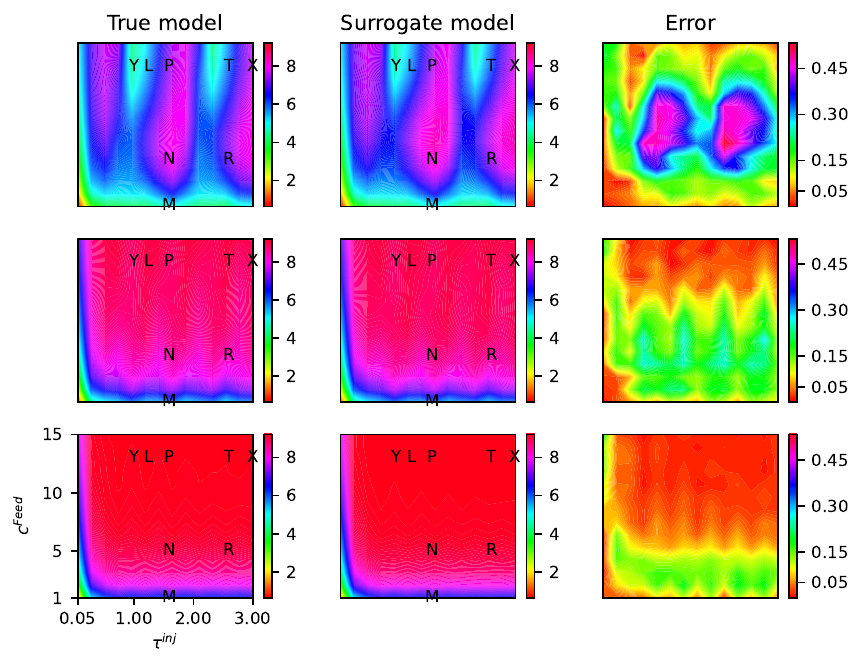}
\caption{Utility functions. Left: true models for  8, 15, and 20 temporal nodes of the sensor respectively. Middle: the utility function with the surrogate model. Right: absolute errors.}
\label{Fig_3}
\end{figure}

In all cases we estimated the utility function with a high-resolution forward mapping and with PSLI for $N =  1105$ training nodes. Every training process was carried out on an equidistant grid over the design space with 14 nodes in each direction. The average times needed to train the surrogate model are given in Table \ref{table:4}. We compare the times invested in training the surrogate model by using the high resolution model. In the case of the utility function. We can see how the training time of the surrogate is one order of magnitude smaller than the brute-force evaluation of the utility function with the real model. The 
%training 
time needed to calculate a parameter chain of length 80 000 with MCMC is  three orders of magnitude smaller than that using  the real model. %It gives a cost-benefit time performance acceptable for both cases.} 
The utility functions are presented in Figure \ref{Fig_3}. 

In Figure \ref{Fig_3}, we observe periodic behavior for $N_s = 8$, which gradually diminishes for higher number of temporal nodes. This suggests that the likelihood effectively captures the dominant components of the model relative to the sensor samples. As expected, fewer sensor samples require precise positioning to optimize information capture.
Furthermore, there is a noticeable increase in the utility function with respect to the concentration $c^{\text{Feed}}$ across all cases. Additionally, the behavior concerning $\tau^{\text{inj}}$ appears to become somewhat independent for $N_s = 15, 20$.

\begin{figure}[htp]
\centering\includegraphics[scale = 1]{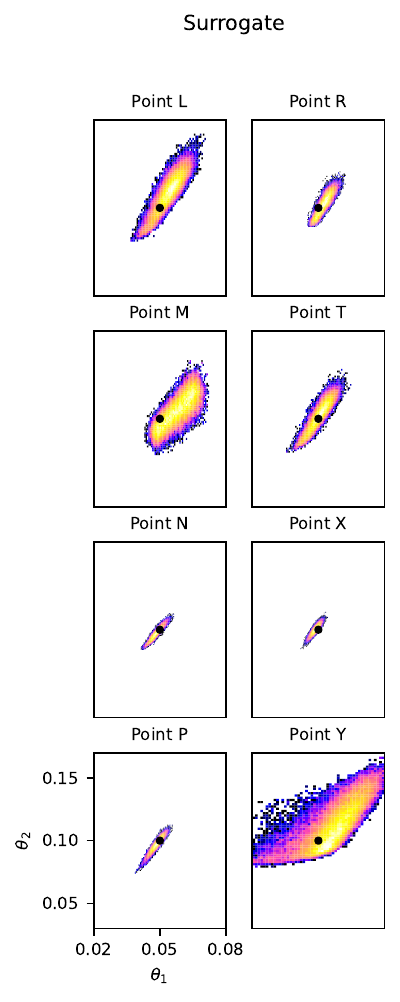}
\centering\includegraphics[scale = 1]{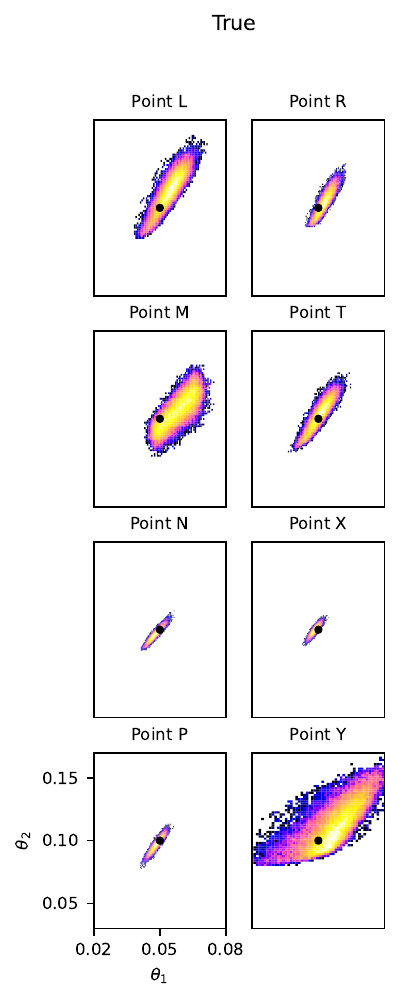}
\caption{Samples of MCMC, $\theta_{1}$ vs $\theta_{2}$. In the left surrogate models for $N_s = 8$. In the right the samples with the true model.}
\label{Fig_99}
\end{figure}

\begin{figure}[htp]
\centering\includegraphics[scale = 1]{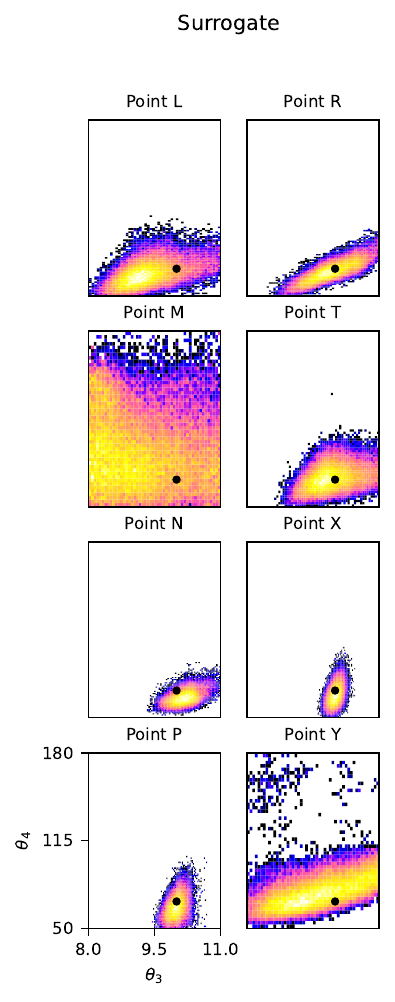}
\centering\includegraphics[scale = 1]{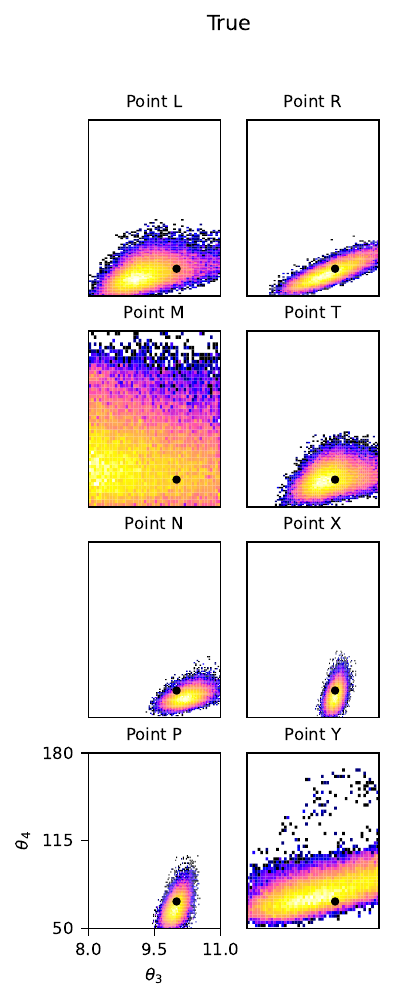}
\caption{Samples of MCMC, $\theta_{3}$ vs $\theta_{4}$. In the left surrogate models for $N_s = 8$. In the right the samples with the true model.}
\label{Fig_99_2}
\end{figure}

\begin{figure}[htp]
\centering\includegraphics[scale = 1]{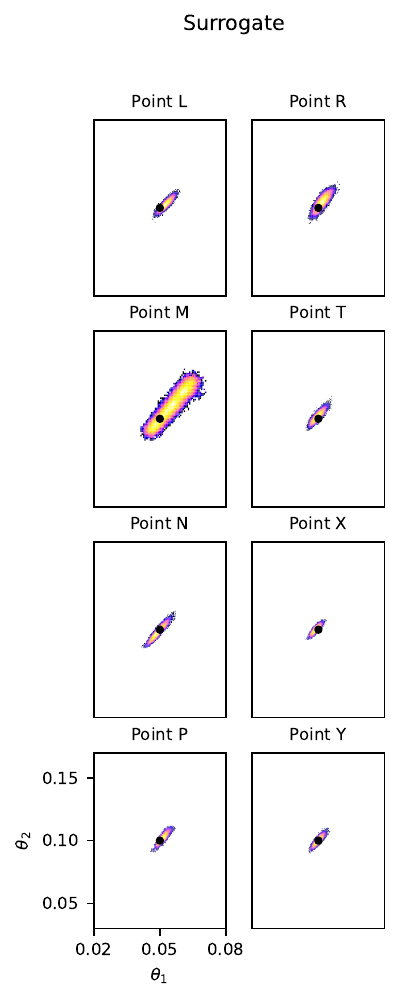}
\centering\includegraphics[scale = 1]{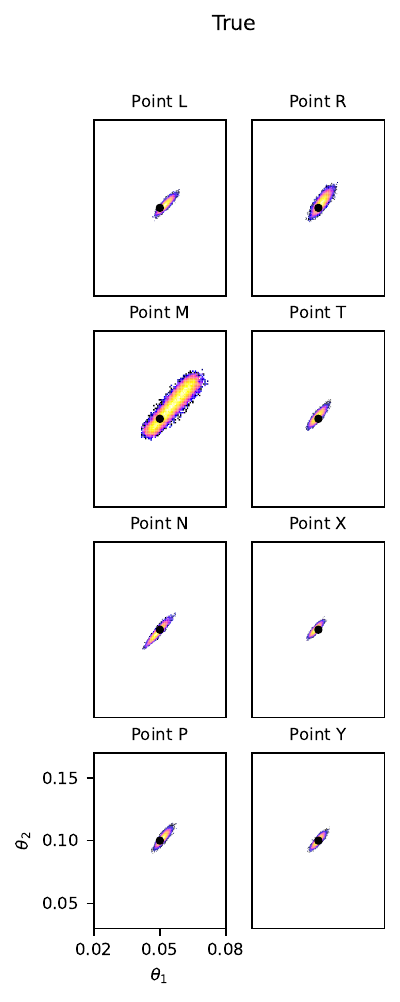}
\caption{Samples of MCMC, $\theta_{1}$ vs $\theta_{2}$. In the left surrogate models for $N_s = 15$. In the right the samples with the true model.}
\label{Fig_100}
\end{figure}

\begin{figure}[htp]
\centering\includegraphics[scale = 1]{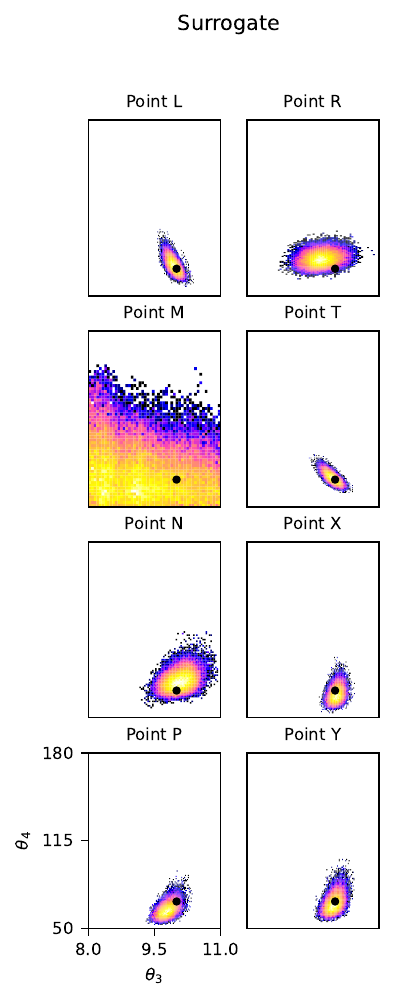}
\centering\includegraphics[scale = 1]{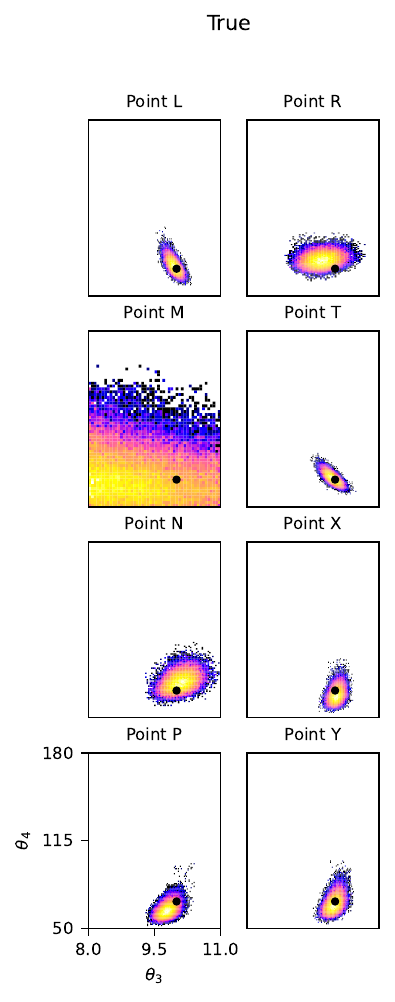}
\caption{Samples of MCMC, $\theta_{3}$ vs $\theta_{4}$. In the left surrogate models for $N_s = 15$. In the right the samples with the true model.}
\label{Fig_100_2}
\end{figure}

\begin{figure}[htp]
\centering\includegraphics[scale = 1]{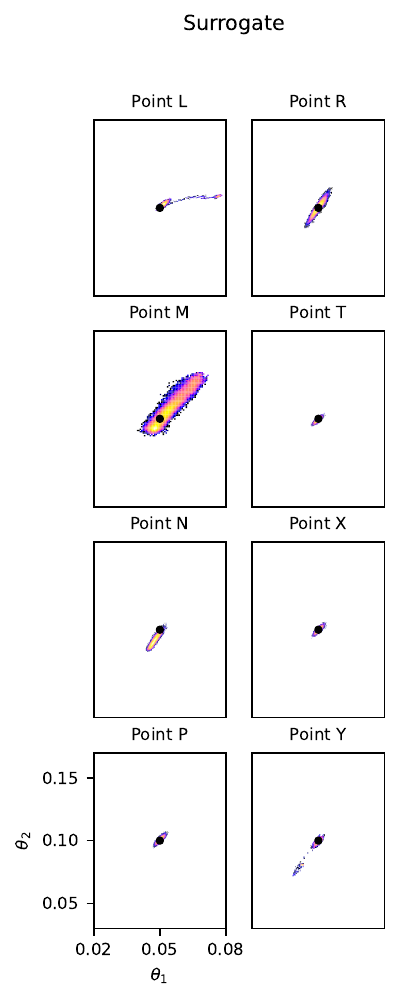}
\centering\includegraphics[scale = 1]{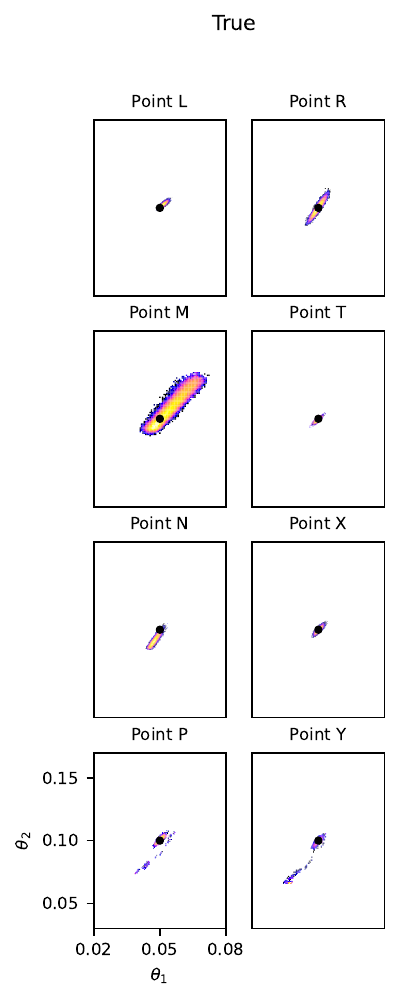}
\caption{Samples of MCMC, $\theta_{1}$ vs $\theta_{2}$. In the left surrogate models for $N_s = 20$. In the right the samples with the true model.}
\label{Fig_101}
\end{figure}

\begin{figure}[htp]
\centering\includegraphics[scale = 1]{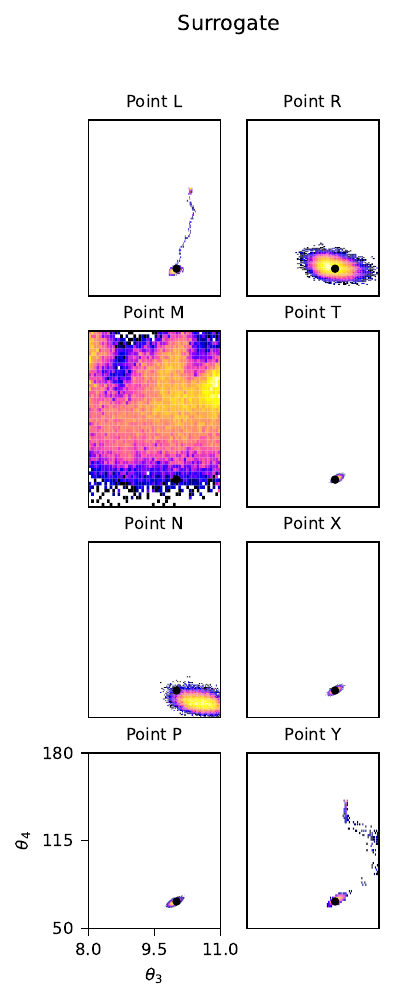}
\centering\includegraphics[scale = 1]{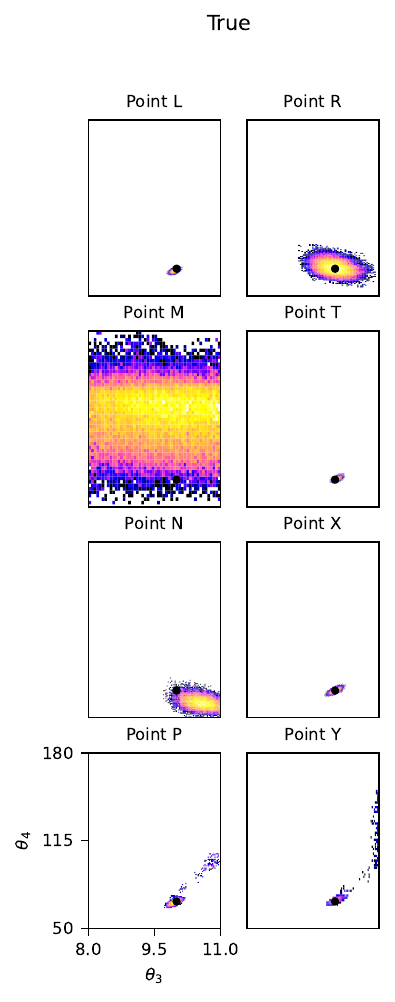}
\caption{Samples of MCMC, $\theta_{3}$ vs $\theta_{4}$. In the left surrogate models for $N_s = 20$. In the right the samples with the true model.}
\label{Fig_101_2}
\end{figure}

To validate the utility function results, we selected 6 design nodes (L, M, N, R, T, and X) and computed MCMC samples using both the true model and the surrogate model at these points. These nodes were chosen to facilitate comparisons across rows and columns, demonstrating that the optimal posterior distribution does not always directly correlate with $\tau^{inj}$ and $c^{Feed}$. The posterior sampling was carried out with the DRAM algorithm \cite{haario2006dram}, comprising 80 000 simulations with the first 30 000 samples discarded to mitigate the burn-in effect. The results are illustrated in Figures \ref{Fig_99}-\ref{Fig_101_2}. 

For $N_s = 8$, we observe periodic behavior in the utility function concerning the time of injection. The posterior distribution with less uncertainty tends to concentrate around point $X$, whereas the distribution with more uncertainty is typically centered at design node $M$. Additionally, we anticipate lower uncertainty in the posterior distribution at point $N$ compared to $L$, and at $R$ compared to $T$.

The periodicity is less pronounced in the case of $N_s = 15$, where we still observe slight oscillations. Additionally, there is a noticeable overall improvement in the samples compared to the previous case. However, we should not anticipate a significant difference in uncertainty between nodes $L$, $T$, and $X$. Interestingly, unlike the previous case, we observe a better distribution at node $T$ compared to $R$.
 
 Finally, for $N_s = 20$, we observe that all nodes except $M$ exhibit similar magnitudes. Once more, there is a notable overall improvement compared to the previous case, with nodes $L, T,$ and $X$ showing similar levels of uncertainty, as well as nodes $N$ and $R$.

\begin{figure}[htp]
\centering\includegraphics[scale = 1]{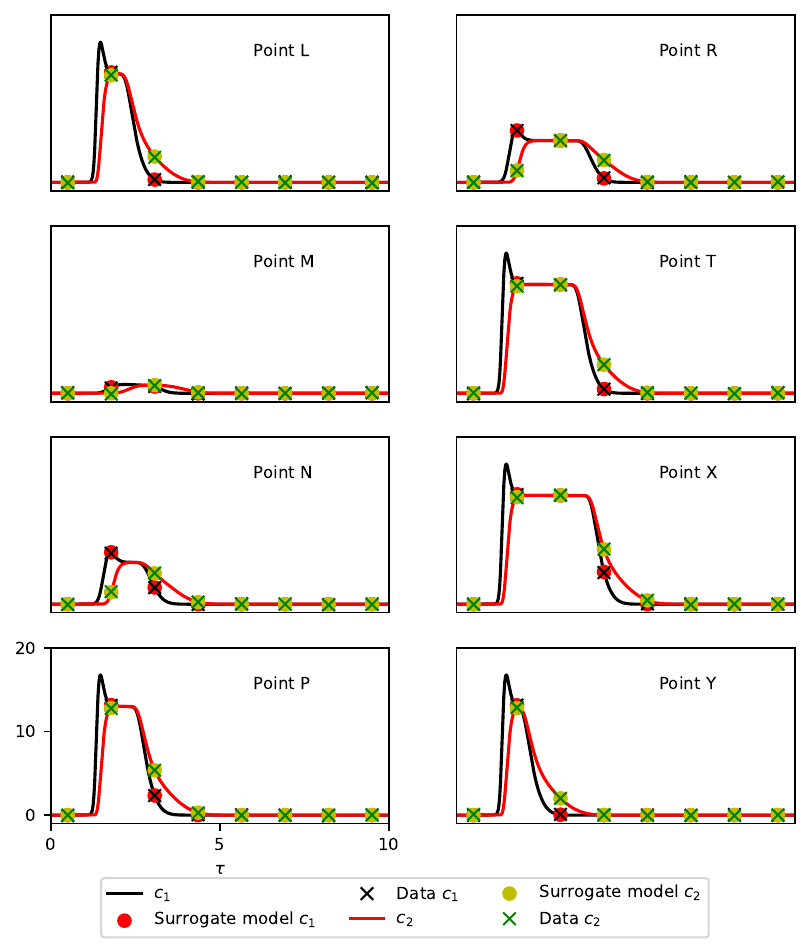}
\caption{Solution to the EDM with $N_s = 8$ over the design points L, M, N, P, R, T, X and Y.}
\label{Fig_96}
\end{figure}

\begin{figure}[htp]
\centering\includegraphics[scale = 1]{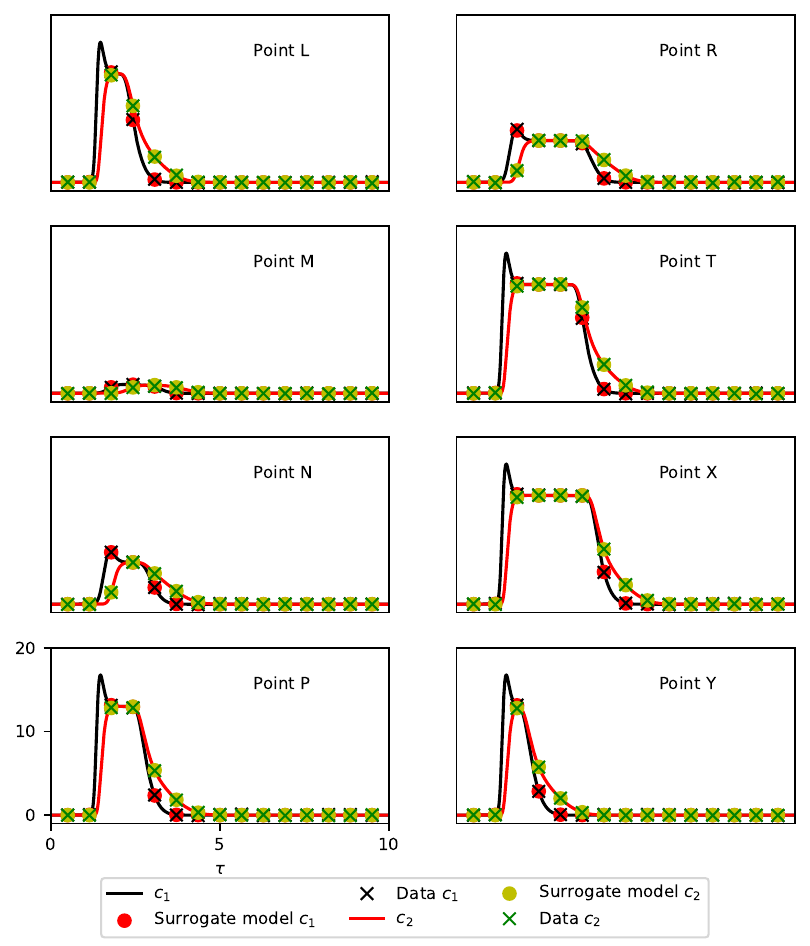}
\caption{Solution to the EDM with $N_s = 15$ over the design points L, M, N, P, R, T, X and Y.}
\label{Fig_97}
\end{figure}

\begin{figure}[htp]
\centering\includegraphics[scale = 1]{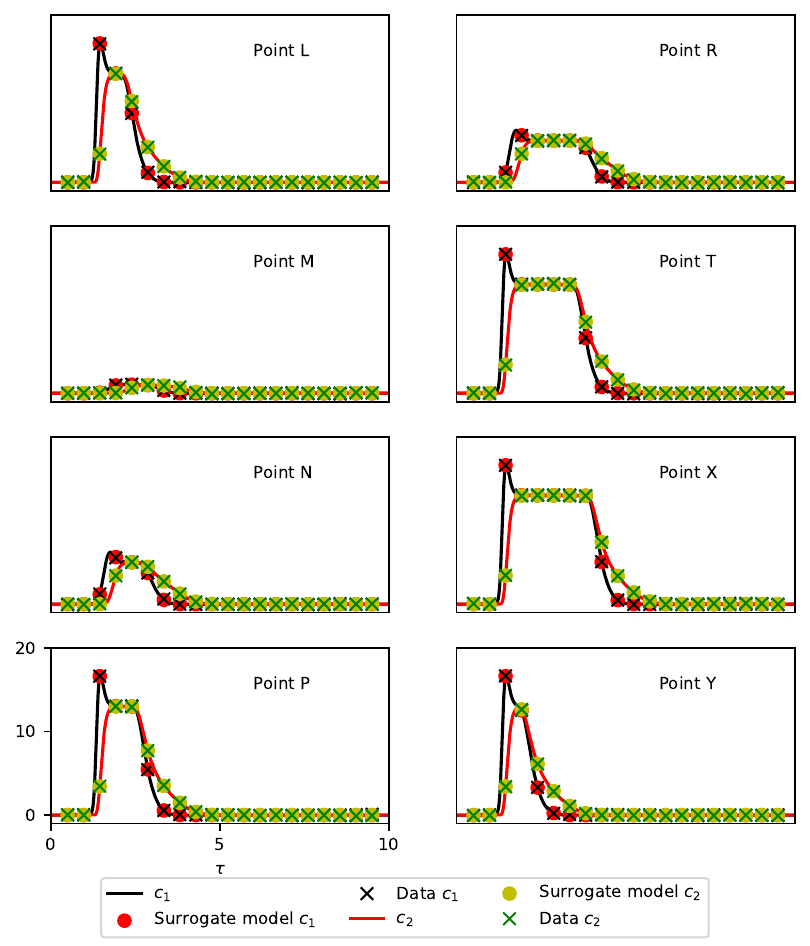}
\caption{Solution to the EDM with $N_s = 20$ over the design points L, M, N, P, R, T, X and Y.}
\label{Fig_98}
\end{figure}

Intuition for the observations above can be drawn from the Figures \ref{Fig_96}, \ref{Fig_97}, and \ref{Fig_98}, where the synthetic data was generated. For $N_s = 8$ and point $L$ we can see how in this narrow peak, only the second temporal node contributes for the component $c_1$. It implies that even though the measurement has larger magnitude than node $N$ or $R$, the node $L$ could not be able to capture the information with perturbations in the parameter space. The same phenomena explains why $X$ and $M$ are the best and the worst design variables respectively.

As we increase the number of temporal observations, our measurements can effectively capture information, even in the presence of potential perturbations in the parameter space. For $N_s = 15, 20$, the dominant factor is $c^{Feed}$; when $\tau^{inj}$ is fixed, the solution structure in the PDE remains consistent, leading to increased peak magnitudes.

% Utilities
%\begin{figure}[htp]
%\centering\includegraphics[scale = 0.29]{Plots/Fig_5_1.pdf}
%\centering\includegraphics[scale = 0.29]{Plots/Fig_5_2.pdf}
%\centering\includegraphics[scale = 0.29]{Plots/Fig_5_3.pdf}
%\centering\includegraphics[scale = 0.29]{Plots/Fig_5_4.pdf}
%\centering\includegraphics[scale = 0.29]{Plots/Fig_5_5.pdf}
%\centering\includegraphics[scale = 0.29]{Plots/Fig_5_6.pdf}
%\centering\includegraphics[scale = 0.29]{Plots/Fig_5_7.pdf}
%\centering\includegraphics[scale = 0.29]{Plots/Fig_5_8.pdf}
%\centering\includegraphics[scale = 0.29]{Plots/Fig_5_9.pdf}

%\caption{Utility functions (time injection $\tau^{inj}$ in $x$ axis and concentration $c^{Feed}$ in $y$ axis). Left: surrogate models for  8, 15, and 20 temporal nodes of the sensor respectively. Middle: the utility function with the true model. Right: absolute errors.}
%\label{Fig_3}
%\end{figure}

% Utilities

%%%%%%%%%%%%%%%%%%%%%%%%%%%%%%%%%%%%%%%%%%%%%%%%%%
% Evaluations
%\begin{figure}[htp]
%\centering\includegraphics[scale = 0.4]{Plots/Fig_2_1.pdf}
%\centering\includegraphics[scale = 0.4]{Plots/Fig_2_2.pdf}
%\centering\includegraphics[scale = 0.4]{Plots/Fig_2_3.pdf}
%\centering\includegraphics[scale = 0.4]{Plots/Fig_2_4.pdf}
%\centering\includegraphics[scale = 0.4]{Plots/Fig_2_5.pdf}
%\centering\includegraphics[scale = 0.4]{Plots/Fig_2_6.pdf}
%\centering\includegraphics[scale = 0.4]{Plots/Fig_2_7.pdf}
%\centering\includegraphics[scale = 0.4]{Plots/Fig_2_8.pdf}
%\caption{Solution to the EDM with $N_s = 8$ over the design points L, M, N, P, R, T, X and Y.}
%\label{Fig_96}
%\end{figure}

%\begin{figure}[htp]
%\centering\includegraphics[scale = 0.4]{Plots/Fig_3_1.pdf}
%\centering\includegraphics[scale = 0.4]{Plots/Fig_3_2.pdf}
%\centering\includegraphics[scale = 0.4]{Plots/Fig_3_3.pdf}
%\centering\includegraphics[scale = 0.4]{Plots/Fig_3_4.pdf}
%\centering\includegraphics[scale = 0.4]{Plots/Fig_3_5.pdf}
%\centering\includegraphics[scale = 0.4]{Plots/Fig_3_6.pdf}
%\centering\includegraphics[scale = 0.4]{Plots/Fig_3_7.pdf}
%\centering\includegraphics[scale = 0.4]{Plots/Fig_3_8.pdf}
%\caption{Solution to the EDM with $N_s = 15$ over the design points L, M, N, P, R, T, X and Y.}
%\label{Fig_97}
%\end{figure}

%\begin{figure}[htp]
%\centering\includegraphics[scale = 0.4]{Plots/Fig_4_1.pdf}
%\centering\includegraphics[scale = 0.4]{Plots/Fig_4_2.pdf}
%\centering\includegraphics[scale = 0.4]{Plots/Fig_4_3.pdf}
%\centering\includegraphics[scale = 0.4]{Plots/Fig_4_4.pdf}
%\centering\includegraphics[scale = 0.4]{Plots/Fig_4_5.pdf}
%\centering\includegraphics[scale = 0.4]{Plots/Fig_4_6.pdf}
%\centering\includegraphics[scale = 0.4]{Plots/Fig_4_7.pdf}
%\centering\includegraphics[scale = 0.4]{Plots/Fig_4_8.pdf}
%\caption{Solution to the EDM with $N_s = 20$ over the design points L, M, N, P, R, T, X and Y.}
%\label{Fig_98}
%\end{figure}

% Samples

%\begin{figure}[htp]
%\centering\includegraphics[scale = 0.45]{Plots/Fig_6_1.pdf}
%\centering\includegraphics[scale = 0.45]{Plots/Fig_6_2.pdf}
%\centering\includegraphics[scale = 0.45]{Plots/Fig_6_3.pdf}
%\centering\includegraphics[scale = 0.45]{Plots/Fig_6_4.pdf}
%\caption{Samples of MCMC. In the left surrogate models for $N_s = 8$. In the right the samples with the true model.}
%\label{Fig_99}
%\end{figure}

%\begin{figure}[htp]
%\centering\includegraphics[scale = 0.45]{Plots/Fig_7_1.pdf}
%\centering\includegraphics[scale = 0.45]{Plots/Fig_7_2.pdf}
%\centering\includegraphics[scale = 0.45]{Plots/Fig_7_3.pdf}
%\centering\includegraphics[scale = 0.45]{Plots/Fig_7_4.pdf}
%\caption{Samples of MCMC. In the left surrogate models for $N_s = 15$. In the right the samples with the true model.}
%\label{Fig_100}
%\end{figure}

%\begin{figure}[htp]
%\centering\includegraphics[scale = 0.45]{Plots/Fig_8_1.pdf}
%\centering\includegraphics[scale = 0.45]{Plots/Fig_8_2.pdf}
%\centering\includegraphics[scale = 0.45]{Plots/Fig_8_3.pdf}
%\centering\includegraphics[scale = 0.45]{Plots/Fig_8_4.pdf}
%\caption{Samples of MCMC. In the left surrogate models for $N_s = 20$. In the right the samples with the true model.}
%\label{Fig_101}
%\end{figure}

\begin{table}[h]
\centering
\begin{tabular}{||c c c||} 
 \hline
  High-res model & Surrogate model & Training \\ [0.5ex] 
 \hline\hline
 $5.6 \times 10^{-1}$ & $1.25 \times 10^{-4}$ & $6.19 \times 10^{2}$  \\ 
$5.6 \times 10^{3}$ & $1.25 \times 10^{0}$ & \\ 
 $4.48 \times 10^{4}$ & $1 \times 10^{1}$ & \\ [1ex] 
 \hline
\end{tabular}

\caption{Average time of evaluations in seconds.  Row one: the forward mapping $G(\theta,d)$, row two: the utility function $U(d)$ with 10000 evaluations of $G(\theta,d)$ at one design point, row three: MCMC parameter chain of length 80 000.}
\label{table:4}
\end{table}

%%%%%%%%%%%%%%%%%%%%%%%%%%%%%%%%%%%%%%%%%%%%%%%%%%
\section{Conclusions}
%%%%%%%%%%%%%%%%%%%%%%%%%%%%%%%%%%%%%%%%%%%%%%%%%%

Our main objective in this study was to develop feasible computational methods for the optimal experimental design in the Equilibrium Dispersive Model (EDM) for chromatography. To achieve this, we implemented a surrogate model and analyzed the utility function with varying numbers of temporal measurement times, aiming to improve both the efficiency and accuracy of parameter estimation.

The surrogate model, based on PSLI, significantly reduced evaluation time compared to solving the PDE system with the Koren scheme. Moreover, it reproduced evaluations from the original model with a small approximation error, as confirmed through simulations of the expected utility functions with different designs. This observation was enforced by considering accuracy in temporal profile prediction and simulation of posterior ensembles to compare concentrations. We note that other methods tested, such as polynomial chaos expansion, failed to produce satisfactory results due to high gradient zones in the concentration profiles.

The analysis of utility function plots revealed that beyond 15 observation time instances there was no further improvement in parameter estimation when varying initial concentrations and injection times, supposing that both design variables exceed a minimal threshold value. 

\section*{Acknowledgement}

This work has been supported by the Research Council of Finland (RCoF) through the Flagship of advanced mathematics for sensing, imaging and modelling, decision number 358 944. Moreover, TH was supported through RCoF decision numbers 353 094 and 348 504. JRRG was supported by the LUT Doctoral School. The authors thank Vesa Kaarnioja for discussions on the PSLI method.

\bibliographystyle{abbrv}
\bibliography{bibliography} 

\end{document}